\documentclass[12pt,preprint]{aastex}

\usepackage{graphicx}

\usepackage{natbib}
\usepackage{rotating}

\newcommand{\be}{\begin{equation}}
\newcommand{\ee}{\end{equation}}
\newcommand{\nn}{\mbox{} \nonumber \\ \mbox{} }
\newcommand{\ba}{\begin{eqnarray}}
\newcommand{\ea}{\end{eqnarray}}
\newcommand{\om}{\omega}

\newcommand{\E}{{\bf E}}
\newcommand{\B}{{\bf B}}

\renewcommand{\v}{{\bf v}}

\newcommand\eg{\textit{e.g. }}

\newcommand{\Bf}{{magnetic field}}

\newcommand{\Ef}{{electric  field}}
\newcommand{\Efs}{{electric fields}}

\newcommand{\NS}{neutron star}
\newcommand{\ms}{magnetosphere}

\newcommand{\Sc}{Schwarzschild}

\begin{document}

\title{High sigma model of pulsar wind nebulae}

\author{Maxim Lyutikov}
\affil{
Department of Physics, Purdue University, 525 Northwestern Avenue
West Lafayette, IN
47907-2036 }

\begin{abstract} 
Pulsars and central engines of long gamma ray burst -- collapsars -- may produce highly magnetized  (Poynting flux dominated) outflows expanding  in a dense surrounding (interstellar medium or stellar material).
For certain injection conditions, the magnetic flux of the wind cannot be accommodated within the cavity. In this case, ideal (non-dissipative) MHD models, similar to the Kennel and Coroniti (1984) model of the Crab nebular,  break down (the so-called $\sigma$  problem). This is typically taken to imply that the wind should become particle-dominated on scales much smaller than the size of the cavity. The wind is then slowed down by a fluid-type (low magnetization) reverse shock.  Recent {\it Fermi} results,  indicating that   synchrotron spectrum of the  Crab nebula extends well beyond the upper limit of the  most efficient radiation reaction-limited acceleration, contradict the presence of a  low sigma  reverse shock.

 We propose an  alternative possibility, that the  excessive magnetic flux is destroyed in a reconnection-like process in two regions:  near the rotational axis  and  near the equator. We construct an example of such highly magnetized wind having two distinct reconnection regions and  suggest that  these reconnection cites are observed  as tori and jets  in pulsar wind nebulae. The model reproduces, qualitatively, the observed morphology of the Crab nebula. In parts of the nebular the dissipation occurs in a relativistically moving wind, alleviating the requirements on the acceleration rate. 
\end{abstract}

\maketitle

\section{Introduction}

The interaction of strongly magnetized, relativistic outflows with a dense surrounding  is a basic problem in pulsar physics, which recently  became  important 
for gamma ray bursts (GRBs)  central engines. This seemingly straightforward problem turns out to be a complicated one, as is illustrated by what became known as the  "$\sigma$ problem"  \citep{kc84}.  Following the work of \cite{reesgunn},  \cite{kc84}
 assumed that the central source produces a wind which blows a cavity in the surrounding  medium. The key assumption of the model is  that in  the bulk of the cavity the wind is  supersonic, causally disconnected from the source.  \cite{kc84} found that for  a non-relativistic expansion  of the bubble,   $v_{\rm out}\ll c$,
   if the Poynting flux in the wind is comparable to or exceeds the particle flux,  there is no acceptable solution for the structure of a  resulting cavity. Introducing a parameter $\sigma $ as  the ratio of   Poynting  $F_{\rm Poynting}$ to particle $F_{\rm p} $  fluxes,    $ \sigma= {F_{\rm Poynting} /F_{\rm p} } $, 
they found that solutions where wind is injected into cavity with relativistic velocities, while  the cavity expands non-relativistically, exist only for $\sigma \sim v_{\rm out}/c\ll 1$. Formally, for $\sigma \geq 1$ the reverse shock in the wind is relativistic with respect to the downstream flow, and since the flow matches to the slowly expanding cavity, the shock is driven back into the pulsar  on a  light crossing time. As a result, the initial set up of the problem, that of a central engine producing relativistic wind, breaks down.  The requirement 
$\sigma \ll 1$ in the bulk of  the cavity runs contrary to  models of pulsar {\ms}s, which  all predict generation of highly magnetized, $\sigma \gg 1$ wind \citep[][]{gj70}. These expectations are confirmed by recent numerical simulations \citep{Spitkovsky06}.
\cite{kc84} concluded that $\sigma $ must change from being much larger  than unity at the light cylinder to much smaller than unity in the bulk of a PWN.  

The conversion of high $\sigma$ to low is very hard to achieve in ideal MHD \citep{HeyvaertsNorman}, though a number of recipes have been proposed \citep[\eg][]{Vlahakis04}.  It is beyond the scope of this paper to discuss this mathematically very difficult  problem, but  simple, non-dissipative  models do not achieve high to low $\sigma$ conversion \citep[the applicability of fluid approach has also been questioned][]{LyubarskyStripedWind}.  

Note, that  the requirement of evolution of $\sigma$ from high to low  follows  from a self-consistency of the model, in a sense that for $\sigma \geq 1$ the whole model, that  of a central pulsar producing a supersonic  wind which is shocked within a PWN,    becomes inconsistent.
Thus, a question still remains, {\it if $\sigma $ remains high, how would a flow behave?}.  This is the main question addressed in this paper.  
 
Recently this problem became important for long GRBs.
Detection 
of Type Ic supernovae nearly coincidence with long GRBs unambiguously linked them
 with deaths of massive stars \citep{smg+03,Hjorth}.
 Studies of the host galaxies 
 of long GRBs, which turned out to be actively star-forming,
 further strengthens this association \citep{Djorgovski}. The outflow
 resulting in a GRB  can be driven by a ''millisecond magnetar'' \citep[a
fast rotating strongly magnetized protoneutron star, \eg][]{Usov92} or a black holes with an accretion disc (e.g. MacFadyen \& Woosley 1999). In both cases the  wind produced by the central engine may be 
  Poynting flux dominated, $\sigma \gg 1$. The evolution of the resulting bubble inside the star then resembles  the evolution of a PWN. Again, in order to achieve a self-consistent model, it  is required that $\sigma$ changes from high to low values \citep{2008MNRAS.385L..28B,2009MNRAS.396.2038B}. 
  (In passing we note that   the electro-magnetic model of GRBs \citep[][]{lb03,LyutikovJPh} explores the possibility that $\sigma$ remains high after the flow breaks out from the star.)
  

\section{Radiation reaction-limited  acceleration of leptons and {\it Fermi} spectrum of Crab}

There is an upper limit on the frequency of synchrotron emission by radiation reaction-limited  acceleration of electrons. If the accelerating \Ef\ is a fraction $\eta \leq 1$ of the \Bf\ (this is equivalent to  acceleration on time scale of inverse cyclotron frequency $1/( \eta \om_{B,rel}) $, where $\om_{B,rel}=\gamma/\om_B$ is relativistic cyclotron frequency of a particle), equating the acceleration rate and synchrotron energy losses,
\be
\eta e B c =  {2 \over 3}  {e^2 \over c}  \gamma^2 \om_B^2
\ee
the peak frequency of emission becomes
\be
\om =  \eta  \om_{\rm max},\, \om_{\rm max}=  0.44 { m c^3 \over e^2}
\ee
This corresponds to energies \citep[see also][]{1996ApJ...457..253D}.
\be
\epsilon_{\rm max} = \hbar \om_{\rm max} =  0.44 \hbar  { m c^3 \over e^2} \approx 30 \mbox{ MeV}.
\label{emax}
\ee
Note, the upper limit (\ref{emax}) assumes {\it non-stochastic}, DC-type acceleration. The break energy observed by {\it Fermi} satellite   in Crab is $\sim 100$ MeV \citep{2009arXiv0909.0862F},  already three times higher than the limit (\ref{emax}). If the 
spectrum above the break requires presence of higher energy particles, those particles should be accelerated by an \Ef\  much larger than the  \Bf.  This is highly problematic under astrophysical conditions, where abundant supply of particles ensures that $E\leq B$. 

There is a number of possible resolutions of this contradiction. First, \Bf\ can be strongly inhomogeneous, spatially or temporarily, so that
particles are accelerated in a low \Bf\ regions, \eg\ very close to the equatorial current sheet, and then emit when they enter a high field region. The possibility of spatially inhomogeneous \Bf\  is also consistent with the results of recent numerical simulations of relativistic shocks, indicating that  only unmagnetized shocks   are efficient accelerators \citep[][another case of highly efficient acceleration,  quasi-parallel  shocks, is not applicable to PWNe]{SironiSpitkovsky09}.
The problem with this scenario  is  that the Crab nebular is a very efficient emitter: about $20\%$ of the spin-down luminosity is converted into radiation. If acceleration is limited to a narrow current layer containing  only a small part of the spin-down luminosity, this put an unreasonable  demands on the efficiency of acceleration. 

Temporal variations in the strength of \Bf\ near the termination shock, seen in numerical simulations by \cite{2009arXiv0907.3647C}, remain is distinct possibility, with a clear prediction: the Crab synchrotron spectrum should be strongly variable  on time scale of month, of the order of wisp variability,  with  the cut-off energy dropping well below 100 MeV. This can be tested with the ongoing observations by {\it Fermi} satellite. Previously, EGRET data did indicate a moderate level of the cut-off energy  variability \citep{1996ApJ...457..253D}.

Another possibility to explain a high acceleration rate, is that the flow is relativistic in the post-shock region. In this case the observed emission is beamed by the  bulk Lorentz  factor $ \Gamma $, 
$\om \sim  \eta \Gamma   \om_{\rm max}$. Thus, a high post-shock Lorentz factors $\Gamma$ may alleviate the problem of super-efficient acceleration.
Relativistic post-shock flows are indeed expected within the \cite{KomissarovLyubarsky} model of PWNs \citep[see also][]{DelZanna04}. In this model, the reverse shock is generally oblique, so that the post-shock flow is moving relativistically. But  the post-shock flow is only mildly relativistic,  independently of the pre-shock Lorentz factor, $\Gamma \sim 1/\theta \sim $ few, where 
$\theta \sim 1$, is the angle that the radially moving  unshocked wind makes with the forward shock (this angle can be small near the axis, but very little energy is coming along those directions). 

As we show below, the {\it  high sigma flows allow dissipation  in a relativistically moving plasma,} boosting the  maximum observed synchrotron frequency to $\om \sim  \eta \Gamma   \om_{\rm max}$ and alleviating the requirements on $\eta$, which can be of the order  $\eta \sim 1/\Gamma \ll 1$.

\section{The $\sigma $ problem}

The $\sigma $ problem of   \cite{kc84}  implies that
 ideal, non-relativistic,  homologous expansion of a bubble of strongly magnetized 
 plasma  injected with nearly a speed of light, 
   cannot occur.
The reason is that magnetic flux and energy are supplied to the 
inflating bubble by the rates that cannot be accommodated in the bubble.
The  rate  of supply is determined by the processes inside the
light cylinder of a compact object, while inflation of the  bubble is controlled by the
external gas density. 
Formally, the surface of the bubble will influence back on the source and would start to control the injection rate, consistent with the allowed rate of expansion. But even if the wind remains subsonic, it is unlikely that processes at the edge of the inflating
bubble would influence  the  wind generation region near the light cylinder due to large separation in scales, thousands in case of GRBs and billions in case of PWNs.

One way to approach the $\sigma$ problem, is to try to solve magnetohydrodynamical reverse  shock conditions in the strongly magnetized wind,  that match the slowly expanding PWN cavity downstream. This can only be done for low $\sigma$: for $\sigma \geq 1 $  the reverse shock is  driven back into the pulsar on light crossing time at which point   the model becomes meaningless \citep{kc84}.


To illustrate the $\sigma$ problem from another viewpoint, assume that the central source  injects  a relativistic wind that carries electromagnetic power and magnetic flux in  a form of toroidal \Bf. The central source injects energy and magnetic flux into a spherical cavity, PWN, with  the injection velocity   $v_{in} =  \beta_{in} c $ and the outer boundary of the cavity  expands with velocity  $v_{out} = c 
\beta_{out} \ll v_{in}$. The core assumption here is that the injection velocity is independent of the conditions in the nebular.
We assume that the expansion velocity of the nebular is much smaller than both the injection velocity   and the fast magnetosonic speed inside a bubble, so that the interior is in  a state of a  quasistatic equilibrium.

Let us assume that magnetic flux is injected into the Crab nebular at some rate $\dot{\Phi}$. This rate depends on the integrated polar angle-dependent luminosity $L(\theta)$ produced by the pulsar. Inside the nebular, where the velocity of expansion becomes strongly sub-fastmagnetosonic, the \Bf\ will relax to a  minimum energy state consistent with a given magnetic flux. It may be shown that such a state corresponds to the current concentrated on the symmetry axis. Since such a current distribution would correspond to the infinite current density, we will assume  that 
there is a current-carrying core of  conical shape with opening angle $\theta_c \ll 1$. Equivalently, the cavity is in a state of dynamical balance and since the pressure forces are negligible,   the  \Bf\ relaxes to a state of  (quasi)static force-free equilibrium with no velocity shear and associated \Efs. The only possible equilibrium of a purely toroidal \Bf\ corresponds to a line current (see,  \eg, Eq. (\ref{main})), with infinite current density on the axis. This state should have the magnetic flux equal the total injected flux.

The central source may have anisotropic energy flux. As an exemplary case,  consider power distribution produced by  an aligned pulsar \cite{michel73}. In this case $L \propto \sin^2 \theta$, and 
normalizing \Bf\ in the wind to the total luminosity $L = 2 \pi  r_{in}^2  \beta_{in} c \int {B^2 \over 4 \pi} \sin \theta d \theta$, the \Bf\ at the inner boundary of the nebular  is 
\be
B = {\sin \theta \over r_{in}} \sqrt{ 3   L \over 2  \beta_{in} c}
\ee
The rate of the injection of the magnetic flux  is 
$ \dot{\Phi} = \beta_{in} c \int B r dr d \theta= 2 \sqrt{ 3 \pi  \beta_{in} L c} $,
so that total  flux stored in a cavity  is
\be
\Phi_{\rm tot}=   \sqrt{ 6 \beta_{in}  L c} \, t
\ee
(and total energy $E=L t$). We assume for simplicity that the luminosity is constant in time.

Inside the nebular, \Bf\ corresponds to a current-carrying core of conical shape with opening angle $\theta_c$.  Equating the total stored flux in such  a configuration to the injected flux, the \Bf\ inside a PWN is 
\be 
B_{in} ={ \sqrt{6 c \beta_{in} L} t \over r \sin \theta R(t)  \ln \tan (\theta_c/2)}
\ee
where $R(t)$ is the outer radius of a PWN. 
The ratio of the  stored  to the injected energy is
\be
{ \int B_{in} ^2/( 8 \pi) dV \over L t} =  {3 \over 2} { \beta_{in} \over  \beta_{out} } {1\over \ln \tan \theta_c/2},
\ee
where we assumed $R(t)= \beta_{out} c t$.
For injection velocity of the order of the speed of light, $\beta_{in} \sim 1$ and expansion velocity several thousand km/sec, $\beta_{out} \sim 10^{-2}$, the angle $\theta_c$ required to accommodate all the injected flux and energy  is unphysical small. Thus, the assumption of the model,  of a source injecting at a  given rate magnetically dominated, ideal  flow into 
slowly expanding cavity, lead to unphysical consequence.

Finally, in  the simplest example of the paradox, consider a cavity of fixed volume, thus neglecting  the expansion completely, The energy and the  magnetic flux are injected into the cavity by the wind linearly in time. But since toroidal flux grows linearly in time, the {\it  stored magnetic energy  grows quadratically}, in clear contradiction to linear injection of energy  \citep{reesgunn}.

\subsection{Possible resolutions of $\sigma$-paradox}

There are several possible resolutions of this    ''paradox''. First, our assumption of high $\sigma$ in the bulk of PWN can be wrong. This is the resolution chosen by  
\cite{kc84}, who postulated that  that $\sigma$ in the wind decreases from $\sigma \gg 1$ to $\sigma \ll 1$.

What if  $\sigma $ remains high in the bulk of the flow? One possibility is that the flow would break out of the cavity: in laboratory devices, where the sizes of the \Bf\ generator and the enclosing cavity are similar, 
and where \Bf\ is limited by strength of cavity's material, 
this sometimes leads to spectacular accidents. In case of GRBs this will occur at the time of jet break-out, but cannot occur in  PWNe.

Another possibility is that the cavity affects the central engine, so that the rate of injection of the magnetic flux adjusts to the rate with which it can be accommodated.
 We consider the possibility unlikely due to large separation of scales. It is unimaginable that a PWN boundary, located at distances of parsecs away from the central source, affects generation of pulsar winds at scales of hundreds of kilometers. In the case of GRB central engines,  the separation of scale is somewhat smaller, only three-four orders of magnitude, from $10^7$ cm (typical size of light cylinder radius of millisecond \NS\ or \Sc\ radius of a black hole) to $10^{11}-10^{12}$ sm (size of a precollapse type Ib/c progenitor), but still there are some three-four orders of magnitude. We assume that {\it   injection  is independent of the expansion and investigate consequences of this assumption.} 
 
Another possible resolution  of the  $\sigma$-paradox requires destruction of magnetic flux, which in turn requires resistivity and reconnection (reconnection in its  original sense, as a process destroying magnetic flux, not necessarily the magnetic energy). 
Thus, 
 dissipation must become important  in the flow. Typically,  reconnection is most efficient in a limited regions of space associated with high electric current concentration, current sheets.
 Current sheets  produced locally, through turbulent motion of plasma, won't destroy the large scale magnetic flux. The destruction of the  the large scale toroidal magnetic flux  can be achieved in two generic locations: near the axis in an O-type reconnection, or near the equatorial plane, where \Bf\ lines of opposite polarity approach each other (assuming, for simplicity, an aligned rotator).

Thus, a flow may be ideal in the bulk and strongly resistive in regions of small measure.
What would a structure of the flow be in this case? As the magnetic flux and necessarily the magnetic energy (and, thus, magnetic pressure) are dissipated, this would create a force misbalance that would push the flow towards the dissipation regions. Thus, a flow structure brining \Bf\ toward dissipation regions - rotational axis and equator - is  set up. This, in turn,  leads to the
pile-up of magnetic field near the dissipation regions and to faster radial expansion in the direction of those regions \citep[akin to  toothpaste tube effect][]{lb03}.
This effect of magnetic pile-up would be most pronounced near the axis
and may be responsible, as we propose, for generation of GRB and pulsar jets.

Previously,
\cite{2003ApJ...591..366K} discussed \Bf\ dissipation in the pulsar wind. In their model,  the  reconnection occurs locally, between different magnetic polarities in a striped wind. 
In this case,  the total injected magnetic flux is zero if averaged over the pulsar's period. In contrast,  we assume that the pulsar rotation axis is nearly aligned with the magnetic axis, so  that the \Bf\ has a given polarity in each hemispheres. 

\section{Magnetized flow within a spherical cavity}

Next we consider an idealized problem of a high-$\sigma$ dissipative flow bounded by a cavity. 
We expect that a flow pattern is set up in which a magnetized wind  is injected at small radii and is deflected at intermediate regions towards the  dissipation regions. 
First, instead of calculating the form of the cavity self-consistently, we assume it to be fixed and then calculated a flow within it. As the simplest example, we consider a spherical cavity, but the following derivation may be repeated for any of the separable coordinates. 
Second, we consider a case of highly magnetized plasma $\sigma \gg 1$ and completely neglect inertial contribution, the so-called force-free approximation \citep[][note, that the relativistic  force-free approximation is different from vacuum, since  massless charges  create electric currents and charge density that  affect the flow dynamics]{Gruzinov99}.
Third, we assume that dissipation is limited to an infinitely small volume (formally of measure zero) concentrated near the axis and the  magnetic equator, while in the bulk the flow is ideal. 
Fourth, we assume that the flow is generated within a volume small compared to the size of the cavity, so that the  \Bf\ is dominated by axially symmetric toroidal field. Neglect of the poloidal \Bf\  is, perhaps, mathematically  the most important simplifying assumption, since it untangles the structure of the poloidal current from the structure of magnetic flux surfaces, the main complication in solving the  Grad-Shafranov equation. 
Finally, we assume that the flow is in a steady state condition, which implies that the velocity of the flow are much larger than the velocity of the cavity expansion and that the pulsar is an aligned rotator. 

Equations of ideal force-free electrodynamics \citep{Gruzinov99} include Maxwell's equations, 
\ba
\label{twomax}
{\partial{\bf E}\over\partial t}&=&\nabla\times{\bf B}-  4 \pi {\bf j}\\
{\partial{\bf B}\over\partial t}&=&-\nabla\times{\bf E}
\label{maxwell}
\ea
and force-balance
\be
\rho{\bf E}+{\bf j}\times{\bf B}=0,
\ee
 which  allows one to relate   the 
current to the electro-magnetic fields
\be
{\bf j}={({\bf E}\times{\bf B})\nabla\cdot{\bf E}+
({\bf B}\cdot\nabla\times{\bf B}-{\bf E}\cdot
\nabla\times{\bf E}){\bf B}\over 4 \pi    B^2}
\label{FF1}
\ee

In the asymptotic domain,
the only
nonzero components of the fields are $B=B_\phi, E_r, E_\theta$ and  
equations
 (\ref{maxwell}) and (\ref{FF1}) give
\ba &&
\partial_t B = -{1\over r} \partial_r( r E_\theta) + {1\over r}  \partial_\theta E_r
\nn &&
\partial_t E_ \theta =  -{1\over r} \partial_r( r B) +
{ E_r  \over B} \left(
{1\over r \sin \theta} \partial_\theta (  \sin \theta E_\theta) +
{1 \over r^2} \partial_r (r^2 E_r) \right)
\nn &&
\partial_t E_r = {1\over r \sin \theta} \partial_\theta (  \sin \theta B)-
{ E_\theta \over B} \left(
 {1\over r \sin \theta} \partial_\theta (  \sin \theta E_\theta)
+{1 \over r^2} \partial_r (r^2 E_r) \right)
\label{ff1}
\ea
In this formulation, the electrical currents arise exclusively  due to charge transport across \Bf\ by electric drift, 
$j_r = \rho  { E_{\theta} \over B}$, $j_\theta =  - \rho  { E_r \over B}$, where $\rho = \nabla\cdot{\bf E}/( 4\pi)$ is charge density. Thus, they are not subject to resistive decay in the bulk
\citep[\eg][]{LyutikovTear}. 
Toroidal axially symmetric \Bf\  is $B_\phi = 2 I /(c r \sin \theta  )$, where $I$ is a current flowing through a magnetic loop at radius $r$ and polar angle $\theta$. 
 In  ideal, time-stationary case  the \Ef\ is poloidal and  irrotational, $E= - \nabla \Phi$. The force-free condition 
 then gives 
\be
2 \nabla I^2 = r^2 \sin ^2 \theta \Delta \Phi  \nabla \Phi
\label{main}
\ee

The neglect of poloidal magnetic flux makes the problem   under-determined, in a sense that for a given toroidal \Bf\ we may chose various distributions of \Ef, the only constraint being that the flow lines do not intersect the given form of the outer boundary. In case of  axial symmetry, the force balance gives 
 one scalar equation (\ref{main}) for 
 two functions, toroidal \Bf\ and 
poloidal electric field, or, equivalently, electric current and charge density. 
(Note, that in absence of a shear, $\Phi=0$, Eq. (\ref{main}) implies $I=$constant, and \Bf\ corresponding to a line current.)

Equation (\ref{main}) should be solved  for a given form of the cavity. If the boundary is impenetrable, 
the velocity of plasma,  
\be
\v = { \E \times \B \over B^2} = {  r \sin \theta ( {\bf e} _\phi \times  \nabla \Phi) \over 2 I}
\label{v}
\ee
should vanish on the boundary. 
Thus, for a fixed spherical  cavity of size $R$, it is required that 
\be
\partial _\theta \Phi(r=R) =0
\label{R}
\ee

\begin{figure}[h!]
   \includegraphics[width=0.95\linewidth]{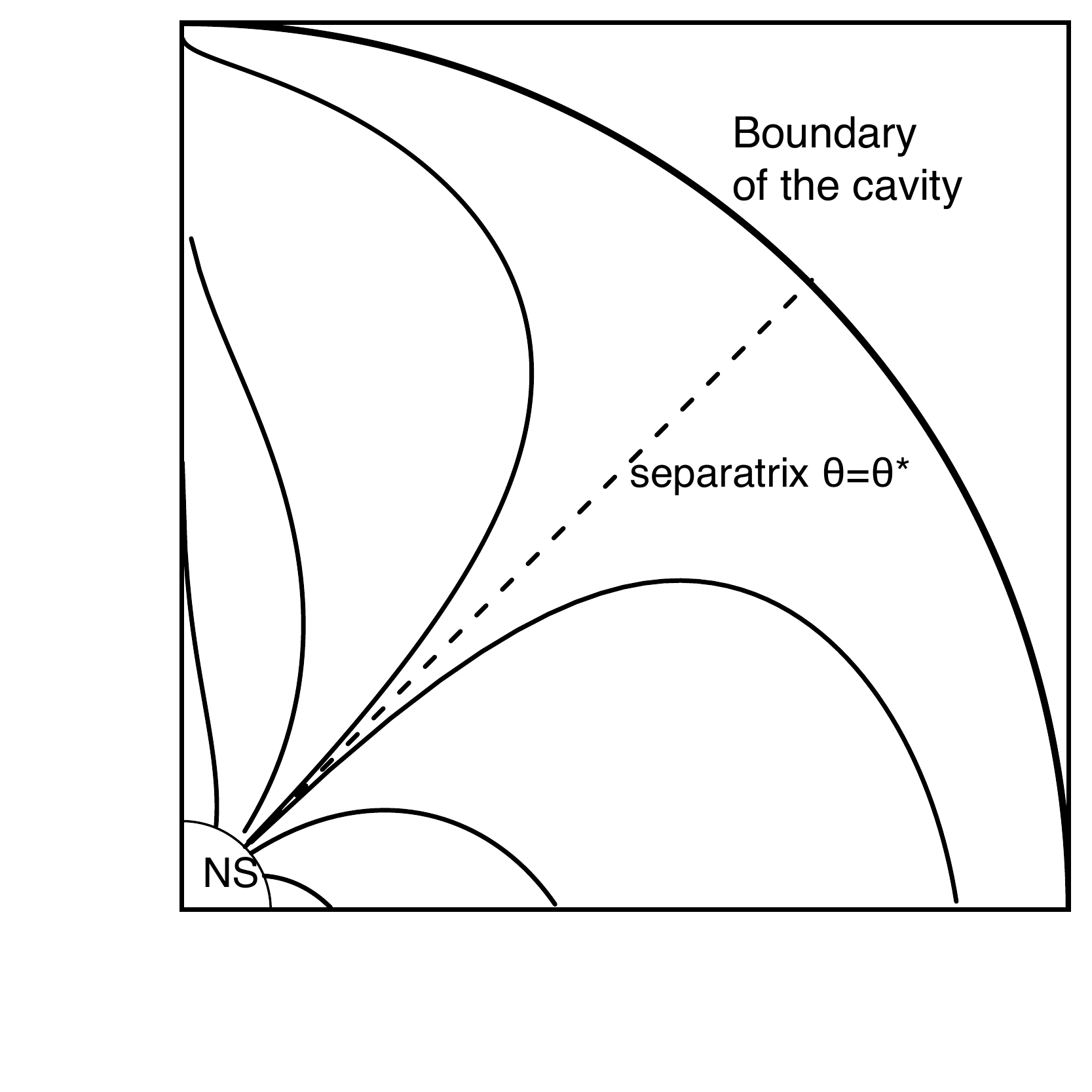}
   \caption{Flow lines given by eq. (\ref{r}). The cavity is assumed to be spherical. Injection of the wind occurs at some small radius $r_0$, labeled NS. 
   Flow lines that star at $\theta < \theta^\ast$ ($45^\circ$ in this example) terminate near the axis, while those that start at $\theta > \theta^\ast$ terminate on equator.
   }
 \label{flow}
 \end{figure}

Equation (\ref{main}) is underdetermined since it  involves two unknown functions. 
A  particularly simple choice of electric potential $\Phi$ and current $I$ corresponds to vanishing charge and current densities in the bulk, so that  $\Delta \Phi =0$ and $I= $constant. 
(Another analytically tractable case is discussed in Appendix \ref{separable}.) Taking the first terms in a series of harmonic functions (Legendre polynomials with $m=0,1$),
 the potential that satisfies boundary condition (\ref{R}) is
\be 
\Phi = \Phi_0 \left( 1- {R \over r}\right) \ln {  \tan {\theta \over 2} \over  \tan {\theta^\ast \over 2}}
\label{Phi}
\ee 
where constant $\theta^\ast $ is the polar angle that separates flow lines that end on the axis, for 
$\theta_ 0 < \theta^\ast$ and at equator, for $\theta_ 0 > \theta^\ast$, where $\theta_0$ is the polar angle at the injection radius $r =r_ 0$. 
The flow lines are then given by (see Fig. \ref{flow}) $
\partial _\theta r = - { \partial _\theta \Phi /\partial _r \Phi}
$, which integrates to
\be
r= { \ln {  \tan {\theta \over 2} \over  \tan {\theta^\ast \over 2}}  R  r_0  \over 
 R \ln {  \tan {\theta_0 \over 2} \over  \tan {\theta^\ast \over 2}} + r_ 0 \ln {  \tan {\theta \over 2} \over  \tan {\theta_0 \over 2}}}
 \label{r}
\ee

The flow lines converge either on the axis or at the equator. 
 It is implicitly assumed that there the dissipation sets in and destroys the \Bf\ at the end  points of flow lines.
On the axis all the flow lines asymptote to the point $\theta =0$, $r=R$, they never cross the line $\theta =0$ at $r<R$. 
Situations at the equator is very different, 
 a flow line that starts at $r_0$, $\theta_0$ reaches equator at 
 \be
r_e = { r_0 R \ln \cot (\theta^\ast/2) \over
r_0 \ln \cot (\theta_0/2) + R \ln  {  \tan {\theta_0 \over 2} \over  \tan {\theta^\ast \over 2}} }.
\ee Thus, magnetic energy must be dissipated continuously and effectively at the equator, while near the axis a toothpaste-like effect pushes the flow in radial direction. Thus, on the axis all the dissipation (formally) happens at  a point  $\theta =0$, $r=R$.

The flow velocity corresponding to Eq. (\ref{Phi}) is 
\ba &&
v_r =\beta_0 \left( {R \over r} -1 \right) 
\nn &&
v_\theta =\beta_0  { R \sin \theta \over r}    \ln {  \tan {\theta \over 2} \over  \tan {\theta^\ast \over 2}} 
\label{vv}
\ea
where we introduced $\beta_0 = {\Phi_0 /  (2 I)}$. 
The flow settles down on the axis, $v_\theta \rightarrow 0$ as $\theta \rightarrow 0$, and has a finite $\theta$-velocity at equator,
  $v_\theta \rightarrow -  \beta_0  \ln   \tan {\theta_0 \over 2} (R/r) $ at $\theta \rightarrow \pi/2$.
  
  The corresponding Poynting flux toward the equator  is
  \ba &&
  F_r = {2  (R-r) \over r^3 \sin^2 \theta} I \Phi_0
  \nn &&
  F_\theta = {2 R \over \sin \theta r^3}   \ln {  \tan {\theta \over 2} \over  \tan {\theta^\ast \over 2}} I \Phi_0
  \ea
  The energy flux towards the reconnection regions is $\propto  r \sin \theta F_\theta$.
Near the axis, the energy  flux diverges as $\ln \theta$; assuming that resistivity becomes important within a  core defined by the polar angle
$\theta_c$, the ratio of   energy  fluxes going to the axis and to  the equator is 
$\xi= \ln \theta_c / \ln \cot (\theta_0/2)$. 
We expect that in case of PWNe, the observed relative brightness of jet and tori depends on the relative energy fluxes to the corresponding reconnection regions, so that the parameter $\xi$ is a measure of the relative brightness within this model. At  a present stage, it is expressed in terms of two formal parameters, the core angle $\theta_c$ and the separatrix angle
$\theta^\ast$. 


To summarize, we found a flow of strongly magnetized plasma within  a given spherical cavity. The particular  model, chosen mostly for simplicity, has a separatrix angle $\theta^\ast$ that delineates the flow lines ending on the equator and on the axis. The ratio of energies dissipated depends also on the core angle $\theta_c$, where dissipation sets in near the axis. The overall value of the velocity scales with the parameter $\beta_0$.
\subsection{Applicability}

So far, the parameters $\beta_0$ and $\theta^\ast$ were arbitrary. We should now check the conditions that  the resulting solutions correspond to a physically meaningful velocity 
\be 
v= \beta_0 \sqrt{ (1-R/r)^2 + (R/r)^2 \ln ^2 { \tan \theta/2 \over  \tan\theta^\ast/2} \sin^2 \theta} < 1
\ee
 at least somewhere inside the sphere $r=R$. Demanding that  $v<1$ on the equator at $r=R$  
 requires that 
\be 
\beta_0  \ln \cot {\theta^\ast \over 2} < 1
\label{param}
\ee
For parameters satisfying Eq. (\ref{param}),
the condition $v< 1$ is violated for sufficiently small radii, $r/R \leq \beta_0  \sqrt{1+ \ln ^2   \cot \theta^\ast/2}$. Thus, the model is not applicable at very  small radii, where mass loading should be important. This is an artifact of our complete neglect of inertia. 

\subsection{Surface brightness of a model PWN}

\begin{figure}[h!]
   \includegraphics[width=0.99\linewidth]{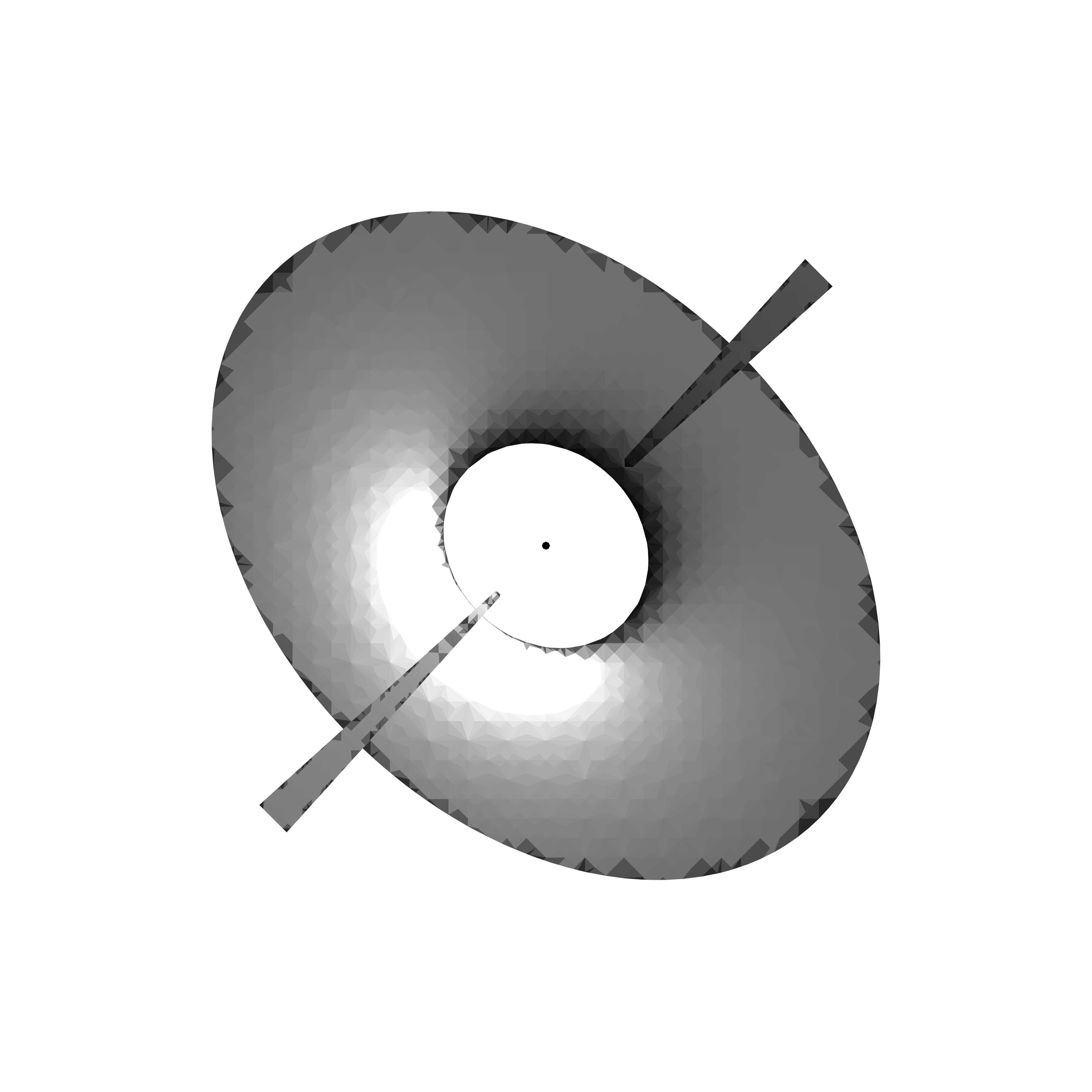}
   \caption{ A negative image of the   Crab nebular in the high-$\sigma$ model. Point in the center denotes the central \NS. Opening angle of the jets is chosen arbitrarily, $0.1$ radian, for illustration purposes. Equatorial and jet emission is proportional to local Doppler factor. Within the central cavity the model is not applicable; see text for  more details.}
 \label{Crab-Fit}
 \end{figure}

To produce an example of the observed PWN in our model, we assume that  (i) the emission is generate in the equatorial plane and along the rotation axis; (2) the observed intensity is proportional to   $\delta ^{2.5}  $, where $\delta= 1/(\gamma (1- {\bf v} \cdot {\bf n}))$ is a  Doppler factor, $\gamma$ is a Lorentz factor corresponding to velocity (\ref{vv}), ${\bf n}$ is a unit vector along the line of sight towards the observer; (3)  chose inclination angle  inferred for Crab nebular, $\theta_{ob} =\pi/3$ (the same value as used by \cite{KomissarovLyubarsky}); (iv) 
chose $\theta^\ast = \pi/3$ (smaller values of $\theta^\ast$ give large Doppler factors and, correspondingly,  more pronounced front-back asymmetry; (v) parameter 
$\beta_0 =0.25$ (the values of $\beta_0 $ mostly  controls the size of the inner region, where the force-free approximation is not applicable; smaller $\beta_0$ correspond to smaller  size of that region); (vi) axis inclination of $135^\circ$ with respect to north-south direction. Results of the calculations are presented in Fig. \ref{Crab-Fit}.

Chandra observation of Crab nebular show that the pulsar is located inside a low brightness bubble, outlined by bright wisps. In our model,  the ``sudden'' onset of dissipation  is an effects of Doppler boosting. 
 Dissipation occurs in a limited region near equator and along the axis. 
 At small radii, when radial velocity is large, the emission is boosted away from the observer. Further out, the flow slows down, as the  radial velocity decreases  with radius $\propto R/r -1$, (\ref{vv}) emission from corresponding parts  comes into view. The resulting structure shows qualitative agreement with observations.
 


\section{Discussion}

In this paper we outlined a model of a high-$\sigma$ (Poynting flux dominated) flow  confined by a medium so  dense, that the rate of the expansion of the blown-up cavity is not sufficient to accommodate the injected toroidal magnetic flux. The model postulates that excessive magnetic flux is destroyed within the nebular at a rate dictated by the expansion of the  nebular. We do not address the related dissipation of energy. Formally, destruction of the flux can proceed at an infinitely small resistivity  \citep[a fast dynamo model,][is an example of model with non-conserved magnetic flux, yet negligible magnetic energy dissipation]{1972SvPhU..15..159V}. 

 The nebular flow in this case is drastically different from the more conventional MHD model of PWNe by  \cite{reesgunn} and \cite{kc84}. The assumption of the model is that if the injection rate of toroidal \Bf\ exceeds the rate at which magnetic flux can be accommodated, a flow will find ways to destroy the excessive magnetic flux. We present an example of how a nebular can achieve this:
 a flow pattern is  set up, where the magnetic energy is injected nearly radially at small distances and is deflected toward reconnection cites in the bulk. 
We give an idealized example of such flows, in which the electric current and electric charge density are confined to two  regions:  near the symmetry (rotational) axis and
equator. In this particular solution  the bulk of the flow is current and charge density-free. This solution is very simple and is meant to illustrate a possible character  of the flow
with the nebular. In spite of it simplicity it reproduces qualitatively two dissipation regions observed in PWNs, axes and tori, as a well as collimated jet-like flows observed in some PWNe.

The model has implications for dynamics of collapsar winds at early stages of GRB explosions.  Even in the current simple version of the model, the flow shows a ``toothpaste tube'' effect: the flow is   collimated towards and is  pushed  along the symmetry axis.  Our assumption of a rigid spherical boundary  precluded formation of a jet. Thus, 
 dissipation on the axis leads to formation of the flow towards the axis and increased \Bf. This provides an additional (to the ideal case) collimation mechanism. 

In addition, we expect that in the equatorial plane the flow also expands faster: conversion of the magnetic field energy density into kinetic energy of particles removes the confining hoop stresses. Thus we expect that   both on the axis and at the equator a PWNs expands faster than at midlatitudes. 

This simple model misses a number of important effects: the properties of the solution at small distances do not quite correspond to what is expected from the  pulsar injection. In the analytical example we discussed the  angular dependence of luminosity does not correspond to what is expected from pulsar models \cite[\eg][]{Spitkovsky06} (the model has more energy near the axis than expected. Also,  at small distances the flow is formally  superluminal.
Dissipation 
of \Bf\ inside the reconnection regions should lead to high plasma pressure, invalidating the assumption of magnetically dominated plasma. In the current formulation, the energy just disappears inside the reconnection regions: this is partly can be attributed to radiative losses, but physically most of the energy just change form, from magnetic to internal and bulk kinetic, and this will affect dynamics close to reconnection regions; the shape of the cavity was assumed to be fixed and given: it should be found self-consistently. Finally, we assumed an aligned rotator, while Crab is, in fact, nearly orthogonal \citep[][]{1999ApJ...522.1046M}. For orthogonal rotators the injected magnetic flux is zero if averaged over the period of rotation, so that the reconnection can proceed locally, not involving large scales flow  discussed in the current model. 

A possible observational distinction between shock acceleration an reconnection is a spectral index of accelerating particles $p$. Shock acceleration typically produces $p \sim 2$ \citep{BlandfordEichler}, where $dn/d\gamma \propto \gamma^{-p}$ \citep[see, though][]{2008ApJ...682.1436L}. Non-linear shock acceleration model can, in principle, produce harder spectra, upto $p=1.5$ for adiabatic index of $4/3$. Observations of Vela pulsar \citep{2004IAUS..218..195K} indicate $p=1.1$, much flatter than expected. In contrast, acceleration in relativistic reconnection is expected to generically produce $p=1$ \citep{2007ApJ...670..702Z}.

How can such a system be simulated on a computer? Due to a relatively limited dynamic range of simulations, the outer boundaries 
will tend to affect strongly the injection region, so that the rate of injected magnetic flux would tend to adjust to the expansion rate. As we argued, this is physically unrealistic. In simulations the (anomalous) resistivity should be adjusted so that all the injected flux (or at least most of it) is destroyed before it is reflected from the outer boundary and affects injection.
   
  Komissarov (2006) gives a numerical example of resistive force-free \ms, which has many similarities to the proposed model. Due to strong  resistivity 
  in the equatorial current, the force-free wind  converges to the equator, so  that all flux surfaces formally become closed. The observed behavior seem to be independent
  of the resistivity, suggesting that a small numerical resistivity is sufficient to relax to a particular solution.   The 
electromagnetic energy flows into the current sheet and disappears inside of it. 

The inertial and kinetic effects will modify the structure of the equatorial current sheet and polar current. Conversion of magnetic energy into heat and plasma pressure will  slow down the influx toward the reconnection regions. On the other hand, increased pressure will lead to faster  radial expansion in those regions and will reduce this effect. Simulations of 
  \cite{Komissarov06}, do show that some flux surfaces close outside the light cylinder in resistive MHD simulations. 
  
 
 The original impetus for this work came from discussions with Roger Blandford. 
 I would like to thank   Elena Amato, Jon Arons, Oleg Kargaltsev, Yuri Lyubarskii, Christopher Reynolds and Mallory Roberts for comments and discussions.
 
\bibliographystyle{apj}
\bibliography{/Users/maximlyutikov/Home/Research/BibTex}

\appendix
\section{Separable solution of Eq. (\ref{main})}
\label{separable}

Eq. (\ref{main}) also allows a separable solution with distributed currents and charge densities
\be
I, \Phi \propto \left( {1\over r^3} -{r^2 \over R^5} \right) \sin^2 \theta
\ee
The radial component of the Poynting flux is $  \propto (r -R^5/r^4)^2 \cos \theta \sin^2 \theta$, with flow lines given by
$
{\sin ^2 \theta \over \sin ^2 \theta_0} = \left( R^5/r_0^3 -r_0^2\right)/(R^5/r^2 -r^3)
$.
All the flow lines  in this solution end at the equator. 
\end{document}